\documentclass[12pt,twoside]{article}   
\usepackage[super,sort,comma]{natbib}

\usepackage{fancyhdr}		

\usepackage[section]{placeins}   %

\usepackage{graphicx}

\usepackage{makecell}
\usepackage{booktabs}
\usepackage{multirow}
\usepackage{amsmath}

\makeatletter \renewcommand\@biblabel[1]{$^{#1}$} \makeatother
\newcommand{\note}[1]{\mbox{}\\ \noindent \rule{16cm}{0.5mm} \\
{\em #1} \\ \noindent \rule{16cm}{0.5mm}
\typeout{    }
\typeout{***********note active on this page *************************}
\typeout{Note: #1  }
\typeout{****************************************end Note}
}

\newcommand{\cen}[1]{\begin{center} #1 \end{center}}

\usepackage[mathlines]{lineno}

\usepackage{hyperref}
\hypersetup{ colorlinks,
    citecolor=blue,
    filecolor=blue,
    linkcolor=blue,
    urlcolor=blue
}

\usepackage{xcolor}

\definecolor{gray}{rgb}{0.6,0.6,0.6}
\definecolor{red}{rgb}{0.85,0,0}
\definecolor{green}{rgb}{0,0.85,0}
\definecolor{blue}{rgb}{0,0,0.85}
\definecolor{beige}{rgb}{0.92,0.87,0.78}
\usepackage[all]{hypcap}    

\begin{document}

\cen{\sf {\small {\bfseries PropNet: Propagating 2D Annotation to 3D Segmentation for Gastric Tumors on CT Scans} \\  
\vspace*{10mm}
Zifan Chen$^{1*}$, Jiazheng Li$^{2*}$, Jie Zhao$^{3}$, Yiting Liu$^{2}$, Hongfeng Li$^{4}$, Bin Dong$^{5{\dag}}$, Lei Tang$^{2{\dag}}$, Li Zhang$^{1{\dag}}$} \\
$^{1}$Center for Data Science, Peking University, Beijing China\\
$^{2}$Key Laboratory of Carcinogenesis and Translational Research (Ministry of Education), Radiology Department, Peking University Cancer Hospital\&Institute, Beijing, China\\
$^{3}$National Engineering Laboratory for Big Data Analysis and Applications, Peking University, Beijing, China\\
$^{4}$Center for Health and Medical Data Science, Peking University, Beijing, China\\
$^{5}$Beijing International Center for Mathematical Research (BICMR), Peking University, Beijing, China
\vspace{5mm}\\
Version typeset \today\\
}

\pagenumbering{roman}
\setcounter{page}{1}
\pagestyle{plain}
$^{*}$ Equal contribution \\
$^{{\dag}}$ Correspondence to Bin Dong: dongbin@math.pku.edu.cn; Lei Tang: tangl@bjcancer.org; Li Zhang: zhangli\_pku@pku.edu.cn 

\begin{abstract}
\noindent {\bf Background:} Accurate segmentation of gastric tumors on 3D CT scans is essential for diagnosis and treatment planning. However, irregular shapes and blurred boundaries of tumors make automatic segmentation challenging. Existing 2D-based methods do not model 3D morphology effectively, and 3D-based methods are computationally expensive and require annotated samples. Some studies have used few-shot learning (FSL) to train models with prior knowledge, but these methods cannot handle irregular shapes and blurred boundaries. \\
{\bf Purpose:} In this pilot study, we anticipate that the human-guided prior knowledge and specially designed modules will address the challenges of 3D segmentation of gastric tumors described above. Drawing inspiration from FSL, we aim to enable the model to learn from information propagation between slices rather than from specific segmentation annotation. \\
{\bf Methods:} We propose the PropNet framework to achieve 3D segmentation of gastric tumors by propagating the radiologist's prior knowledge from the 2D annotated slice into the entire 3D CT space. The proposing stage uses an inter-slice context module to extract contextual information between adjacent slices and generate a coarse segmentation. The refining stage then corrects difficult-to-segment pixels for improved segmentation. We use two-way branches for boundary and region in each stage to enhance the segmentation performance at tumor boundaries. Furthermore, we adopt an up-down parallel propagation strategy to speed up our model's predictions. \\
{\bf Results:} Our study utilizes 98 patients (173 CT scans) as a training set and 30 patients (59 CT scans) as a validation set. The proposed method achieves a comparable agreement with manual annotation (Dice of 0.803 with a p-value of 0.563) and radiologists (Dice of 0.797) on the validation set while significantly improving efficiency with only 40 seconds for annotation and segmentation of a tumor. Furthermore, we test our proposed method on an independent external validation set of 42 patients with stage IV gastric cancer for prognostic verification. The 3D segmentation generated by our proposed method improved prognostic prediction performance (C-index of 0.620 vs. 0.576 for 2D annotations). We conduct experiments to verify stability, including simulating actual usage scenarios and using different radiologists' guidance annotations, all showing comparable performance (Dice between 0.785 and 0.803). Ablation studies confirm the importance of the refining stage and auxiliary boundary branches for irregular shapes and blurred boundaries. \\
{\bf Conclusions:} Our radiologist-guided propagation model efficiently generates accurate gastric tumor segmentation by leveraging specially designed modules and radiologist-guided slices. The method exhibits stability in simulated medical scenarios and improves prognostic performance for survival analysis. Our proposed model provides a solution to accelerate the quantitative analysis of gastric tumors, reducing the workload of high-throughput image reading and improving downstream task performance. \\

\end{abstract}

\note{Key words: gastric tumor segmentation, deep learning, computed tomography (CT)}

\newpage     

\tableofcontents

\newpage

\setlength{\baselineskip}{0.7cm}      

\pagenumbering{arabic}
\setcounter{page}{1}
\pagestyle{fancy}
\section{Introduction}
\label{sec:introduction}
Gastric cancer is one of the most common cancers and the leading cause of worldwide cancer-related death~\cite{jemal2011global,van2016gastric,takahashi2013gastric,yada2013current,sung2021global}. Biopsies are crucial for diagnosing and determining molecular subtypes, but the intra-heterogeneity of gastric cancer may limit their value~\cite{park2020pd}. Computed tomography (CT) scans provide a non-invasive way to comprehensively assess gastric cancer, offering information for treatment, operation planning, response evaluation, and prognosis prediction~\cite{9448400,beer2006adenocarcinomas,lu2013consideration,wang2017ct}. However, manually segmenting gastric tumors on CT scans is time-consuming and impractical for radiologists~\cite{beer2006adenocarcinomas,xu2018ct}. With recent developments in deep learning~\cite{lecun2015deep, de2018methodology}, automatic recognition of medical images has become a promising tool to solve the segmentation problem. Deep learning has shown superior accuracy compared to experts in automatically classifying lung cancer and melanoma through training on previous CTs and dermatoscopic images, respectively~\cite{ardila2019end, haenssle2018man}. Deep learning can extract massive complex information from medical images and detect implicit patterns, potentially providing a more effective workflow in imaging diagnosis.

Automated image analyses of gastric tumors have traditionally relied on 2D pathological slices or 2D endoscopy images~\cite{xu2019review, e2020deep}. However, the segmentation of gastric tumors on 3D CT scans is more challenging compared to other organs or tumors with regular shapes, such as liver or lung tumors. This is due to several reasons. Firstly, gastric cancer is often characterized by thickening the stomach wall with a curved course, leading to an irregular shape. Secondly, there are many adjacent organs around the stomach with a bit of fat between them. As a result, the transition from gastric cancer to the normal stomach wall is often unclear, leading to blurred boundaries under the relatively low resolution of CT scans.

Recent works have attempted to address this challenge using supervised deep learning-based 3D automated segmentation methods for gastric tumors on CT scans~\cite{zhang20213d,li20213d}. However, these methods require substantial computation and annotated data, which limits their deployment and application in actual clinical scenarios. Few-shot learning (FSL) provides a feasible solution to boost the model's performance by training the model with transferable prior knowledge based on 2D-based architectures using scarce labeled data. For example, SE-Net~\cite{roy2020squeeze} introduces few-shot learning to segment multiple organs using only 85 contrast-enhanced CT scans.

To overcome the limitations of fully supervised segmentation and leverage the benefits of few-shot learning, we propose a human-guided propagating framework, PropNet, for semi-automated gastric tumor segmentation. The framework randomly selects pairs of labeled transverse slices from the same CT to create support-query tasks, each composed of a support slice and a query slice. The proposed PropNet is trained on these support-query tasks, focusing on the propagation relationship between two transverse slices, rather than extracting 3D features. This approach enables the model to learn more general knowledge about how to propagate a single slice to the entire 3D segmentation of gastric tumors with different sizes, shapes, and textures on CT images. Consequently, for a test 3D CT, only a single support slice needs to be annotated to infer the trained model, reducing the human annotation workload from a total 3D volume to a 2D transverse slice.

In addition, we have implemented several improvements to the baseline model to enhance its performance. Firstly, we recognize that the irregular and blurred boundaries of gastric tumors in CT images pose a challenge for accurate segmentation. To address this, we divide the model into two expert branches for boundary and region, prioritizing the prediction of tumor boundaries. Secondly, we introduce a refining stage to minimize possible errors in the segmentation of difficult-to-segment pixels. Finally, to optimize efficiency, we propose propagating the segmentation from the annotated support slice both upwards and downwards, generating a segmentation of the entire tumor in parallel. We conduct extensive experiments and ablation tests to evaluate the effectiveness of our proposed framework using real-world data provided by Beijing Cancer Hospital.

In conclusion, the proposed PropNet framework achieves comparable accuracy to radiologists' manual annotation and outperforms currently available methods for semi-automatic segmentation of gastric tumors on CT scans. The model is efficient, with a manual annotation time of about 40 seconds for a query slice and an additional automated inference time of less than one second on a standard GPU or about 70 seconds on a regular CPU. The effectiveness of the refining stage and boundary-region two-way architecture is demonstrated through ablation experiments. Additionally, an independent external validation set of gastric cancer immunotherapy patients is used to evaluate the model's aid to downstream tasks, such as survival analysis. The proposed PropNet framework has the potential to improve the reliability and quality of diagnosis and reduce the workload of radiologists, making it a valuable assistant tool for medical image segmentation~\cite{robertson2018digital}.

\section{Related work}
\subsection{Deep learning in gastric cancer image analysis}
Deep learning has many applications in gastric cancer detection, classification, and segmentation of 2D pathological slices and 2D endoscopy images~\cite{9107452,8394987,sakai2018automatic,liu2018transfer,hirasawa2018application,zhang2017gastric,sun2018novel,fang2019predicting,li2018gt,liang2018weakly}. For example, deep convolutional neural networks (CNNs) are widely used for screening early gastric cancer lesions on endoscopic images. Researchers have applied several popular CNN backbones and their variants, such as FCN, VGG, Inception, and Inception-ResNet, to the tasks of gastric cancer detection~\cite{sakai2018automatic,liu2018transfer,hirasawa2018application}.

Additionally, deep learning is used for gastric cancer classification, including distinguishing disease types~\cite{zhang2017gastric}, determining benign and malignant lesions~\cite{sun2018novel}, and measuring the degree of differentiation~\cite{fang2019predicting}. Gastric cancer segmentation is also applied to build an end-to-end prediction model from image to label. Clinical knowledge about gastric cancer lesions is usually used to optimize the network model architecture and perform post-processing. For instance, Li et al. reported a multi-branch FCN (GT-Net) to realize gastric cancer segmentation on pathological slices~\cite{li2018gt}, while Qin et al. introduced a new inception structure based on ResNet to utilize multi-scale information for better segmentation~\cite{qin2018large}. Using weakly supervised learning, Liang et al. used the Patch-based FCN to achieve gastric cancer segmentation at a fine-grained level~\cite{liang2018weakly}.

In recent years, a few studies have explored 3D automatic segmentation of gastric tumors on CT scans. Zhang et al. constructed a multi-task model for gastric tumor segmentation and lymph node classification and added a multi-attention module into the 3D structure to improve the model's performance~\cite{zhang20213d}. Li et al. improved the automatic segmentation of gastric tumors based on an improved feature pyramid network~\cite{li20213d}. However, due to the limited voxel-level annotations of 3D gastric tumors, the segmentation accuracy of these 3D automated methods is not high.

\subsection{Few-shot learning}
Recent research on few-shot learning can be divided into three categories. The first category involves learning prior knowledge from a few training samples to select model parameters that can be well generalized. This includes techniques such as initial parameter selection for efficient fine-tuning~\cite{finn2017model, nichol2018first, ravi2016optimization}, and hyper-parameter optimization during training~\cite{bello2017neural, santoro2016one}. The second category involves using generation methods to improve the distribution of the training data~\cite{lake2015human, edwards2016towards}. The third category uses metric learning to determine feature-level similarities between few-shot support samples and queries~\cite{koch2015siamese, santoro2017simple, bertinetto2016learning, snell2017prototypical, sung2018learning}.

Given that obtaining pixel-level annotations is more challenging than obtaining labels in other vision tasks, it is natural to consider applying the few-shot learning paradigm to semantic segmentation. Most previous works propose using a conditional branch to assist the main segmentation branch in achieving few-shot segmentation. For example, Shaban et al.~\cite{shaban2017one} proposed using the conditional branch to encode the support set and modulate the last layer in the segmentation branch. Other studies have used similar conditional branch schemes to impose modulation on the segmentation branch~\cite{rakelly2018conditional, zhang2020sg, dong2018few, zhang2019canet}. On the other hand, Li et al.~\cite{li2020fss} used a relation network to fit the hidden relationship between the support images and the query images, which enhances the segmentation model's generality.

Few-shot learning has also been used to segment regular organs in recent years. For example, Roy et al.~\cite{roy2020squeeze} used a squeeze-and-excitation network (SE-Net) for volumetric segmentation and demonstrated the use of few-shot learning with limited annotations. Sun et al. designed a global correlation network with discriminative embedding to utilize the prior relationship between multiple organs in medical images based on few-shot learning~\cite{sun2022few}. However, these methods focus on regular organ segmentation and may fail to segment gastric tumors with irregular shapes and blurred boundaries. In this work, we aim to overcome these limitations and propose a propagating segmentation framework suitable for segmenting 3D gastric tumors on CT scans.

\begin{figure}[t]
	\centering
	\includegraphics[width=1.0\linewidth]{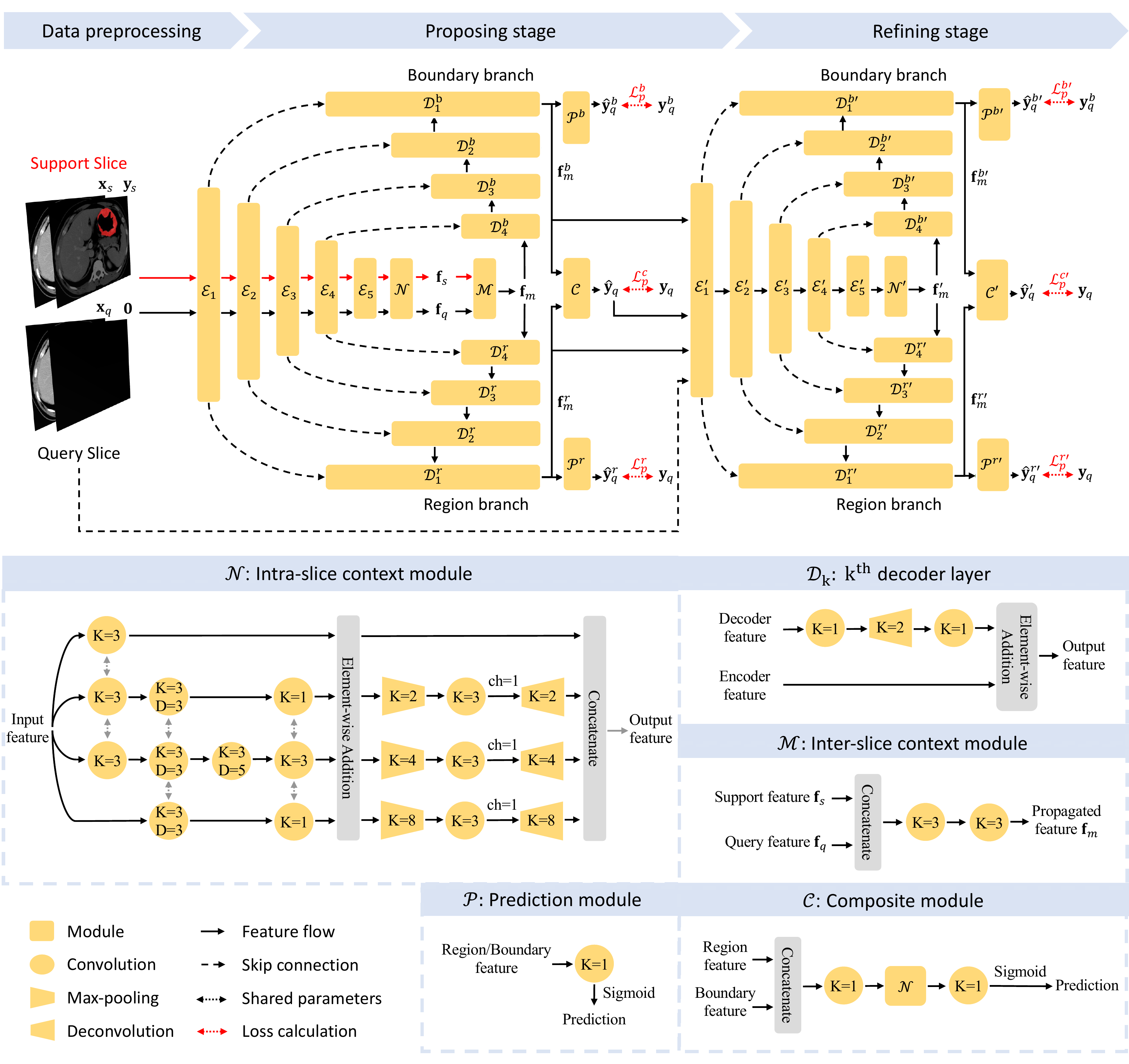}
	\caption{The overview of PropNet. The model takes an annotated slice as the support slice while other slices as query slices as inputs during training. PropNet consists of two stages: 1) the proposing stage uses the inter-slice context module to mine the contextual information between slices for coarse segmentation; 2) the refining stage, which corrects difficult-to-segment pixels predictions. Additionally, a separated boundary branch is introduced into each stage to enhance predictions of blurry boundaries. "K" refers to the kernel size in convolutional, max-pooling, or deconvolutional layers. "D" represents the dilated convolutional stride. "ch" indicates the specified channel number.}
	\label{fig:fig1_overview}
\end{figure}

\section{Methods}

In this section, we will first provide an overview of the proposed PropNet. Then, we will describe the module details for PropNet and introduce the objectives during training. Finally, we will show the model inference using an up-down parallel iteration strategy for test CT scans.

\subsection{Overview}
\label{sec:method_overview}

Figure~\ref{fig:fig1_overview} presents an overview of the proposed PropNet framework. The first step involves the radiologist selecting and annotating a slice, ideally with the largest cross-sectional area of the tumor, from a series of 2D transverse slices $\{\mathbf{x}_1, \mathbf{x}_2, …, \mathbf{x}_Z\}$ in the 3D volumetric CT scan, where $Z$ is the number of the transverse slices. This selected slice is defined as the support slice, denoted as $\mathbf{x}_s$, and its corresponding annotation is $\mathbf{y}_s$. The remaining slices are defined as query slices, and zero maps are used as placeholders to omit annotation information while retaining input sizes. The query inputs are then denoted as $(\mathbf{x}_q, \mathbf{0})$, where $q \neq s$. The goal of PropNet is to expand the support slice annotation into query slices and obtain complete 3D gastric tumor segmentation.

The proposed PropNet comprises two stages: the proposing stage and the refining stage. In the proposing stage, we aim to discover the relationship between the labeled support image and the unlabeled query images. A light-weighted inter-slice context module is implemented to fuse the features from both support and query images, enabling the model to provide a coarse segmentation of the query images. In addition, PropNet segments both the tumor and its boundary area in parallel to enhance the segmentation of the unclear borders of gastric tumors in CT scans. In the refining stage, the model inherits most of the modules in the proposing stage and takes comprehensive input by combining the original query image with the region and boundary logits from the last layer of the decoders in the previous stage. To improve the segmentation results of challenging cases, a difficult-to-segment pixel mining approach is employed to supervise the refining stage.

\subsection{Module Details}
The proposing stage of the model aims to learn the relationship between the support slice $(\mathbf{x}_s, \mathbf{y}_s)$ and the query slice $(\mathbf{x}_q, \mathbf{0})$. Based on this relationship, the model predicts the segmentation $\hat{\mathbf{y}}_q$ of the gastric tumor on $\mathbf{x}_q$. The model uses an encoder-decoder structure based on the classical medical image segmentation method UNet~\cite{ronneberger2015u}.

\noindent\textbf{Feature Extraction:} To balance the network's ability and complexity, we adopt ResNet-34~\cite{he2016deep} as the network backbone for the encoder $\mathcal E$. In Figure~\ref{fig:fig1_overview}, $\mathcal E_1$ to $\mathcal E_5$ represent the first five layers of ResNet-34.

\noindent\textbf{Intra-slice context module:} To encourage the proposed model to capture multi-scale contextual information of tumors within a slice, we introduce an intra-slice context module $\mathcal N$, as shown in Figure~\ref{fig:fig1_overview}. The module contains two multi-scale context-aware components in series. The first processes the features from the encoder $\mathcal E$ in four parallel paths with different receptive fields controlled by dilation rates. It is worth noting that convolutional layers with the same dilation rate in different paths share weights to reduce computation and reuse knowledge. An element-wise addition is then used to fuse the multi-scale information. The second component is in a Pyramid-Scene-Parsing (PSP) style~\cite{zhao2017pyramid}. Features from different PSP branches are concatenated to form the final context-enhanced features within the slice. Based on $\mathcal E$ and $\mathcal N$, the model generates context-enhanced features from the support and query slice, respectively. The corresponding notions are $\mathbf{f}_s=\mathcal N \circ \mathcal E(\mathbf{x}_s)$ and $\mathbf{f}_q=\mathcal N \circ \mathcal E(\mathbf{x}_q)$.

\noindent\textbf{Inter-slice context module:} The inter-slice context module $\mathcal M$, shown in Figure~\ref{fig:fig1_overview}, consists of two $3\times 3$ convolutional layers that produce a propagated feature map containing relationship information between $\mathbf{f}_s$ and $\mathbf{f}_q$. This propagated feature map is denoted as $\mathbf{f}_m$, and it efficiently captures the propagation relationship between the features from the support and query slices by utilizing the contextual information provided by $\mathcal N$. 

\noindent\textbf{Region/Boundary decoders:} The propagated feature map $\mathbf{f}_m$ is fed into two decoders, $\mathcal D^{r}$ and $\mathcal D^{b}$, which extract the tumor's region and boundary features, denoted as $\mathbf{f}_m^r=\mathcal D^{r}(\mathbf{f}_m)$ and $\mathbf{f}_m^b=\mathcal D^{b}(\mathbf{f}_m)$, respectively. Then, two prediction layers are used to generate segmentation, $\hat{\mathbf{y}}_q^{r}=\mathcal P^r(\mathbf{f}_m^r)$ and $\hat{\mathbf{y}}_q^{b}=\mathcal P^b(\mathbf{f}_m^b)$. We follow UNet~\cite{ronneberger2015u} to add skip connections between the corresponding layers of the encoder and the decoders to ensure sufficient shallow features are passed to maintain fine details in the segmentation. As shown in Figure~\ref{fig:fig1_overview}, the low-resolution feature in the decoder passes through a $2\times 2$ deconvolutional layer sandwiched between two $1\times 1$ convolutional layers for up-sampling. Then, it is added with the shallow feature from the encoder. Finally, we use a composite module $\mathcal C$ to fuse the region and boundary branches' features and predict a more accurate result $\hat{\mathbf{y}}_q^c=\mathcal C\left (\mathbf{f}_m^r, \mathbf{f}_m^b\right)$.

\noindent\textbf{Dealing with difficult-to-segment pixels:} The refining stage is designed to improve the coarse segmentation from the proposing stage. It takes $\mathbf{x}_q$ and the features ($\mathbf{f}_m^r$, $\mathbf{f}_m^b$) as inputs. Most modules in the proposing stage are inherited, except for the inter-slice context module. Specifically, the refining stage consists of an encoder $\mathcal E'$, an intra-slice context module $\mathcal N'$, two decoders $\mathcal D^{r}{}'$ and $\mathcal D^{b}{}'$, two prediction layers $\mathcal P^{r}{}'$ and $\mathcal P^{b}{}'$, which separately predict $\hat{\mathbf{y}}_q^{r}{}'$ and $\hat{\mathbf{y}}_q^{b}{}'$, and a final composite module $\mathcal C'$, which produces the improved prediction $\hat{\mathbf{y}}_q^{c}{}'$.

\subsection{Objective Function}
Given the ground truth of tumor segmentation on the query slice $\mathbf{y}_q$, we can obtain its boundary $\mathbf{y}_q^b$ using erosion in morphological operations, as follows:
\begin{align}
    \mathbf{y}_q^b=\mathbf{y}_q\ominus {\rm e}(\mathbf{y}_q,E),
    \label{equ:erode}
\end{align}
where ${\rm e}(\cdot)$ denotes erosion and $E$ represents the kernel size used in erosion.

The losses for the proposing stage's predictions, $\hat{\mathbf{y}}_q^c$, $\hat{\mathbf{y}}_q^r$, and $\hat{\mathbf{y}}_q^b$, can be calculated as $\mathcal L_q^c=\rm d(\hat{\mathbf{y}}_q^c,\mathbf{y}_q)$, $\mathcal L_q^r=\rm d(\hat{\mathbf{y}}_q^r,\mathbf{y}_q)$, and $\mathcal L_q^b=\rm d(\hat{\mathbf{y}}_q^b, \mathbf{y}_q^b)$, where $\rm d(\cdot,\cdot)$ is the soft Dice loss, denoted as:
\begin{align}
    \rm d(\mathbf{a},\mathbf{b})=1.0-\frac{2\left | {\mathbf{a}} \odot {\mathbf{b}} \right |}{\left | {\mathbf{a}} \right | + \left | {\mathbf{b}} \right |}.
    \label{equ:softdiceloss}
\end{align}

For the refining stage, we adopt online difficult-to-segment pixel mining inspired by OHEM~\cite{shrivastava2016training}. We define difficult-to-segment pixels as those whose gradient is in the larger one-third of the current loss map. We use $\mathbf{h}\in{0,1}$ to denote the location of difficult-to-segment pixels, where ones are for difficult-to-segment pixels and zeros are for remaining pixels. Equation~\ref{equ:softdiceloss} can be rewritten as a version for difficult-to-segment pixel mining:
\begin{align}
    \rm d'(\mathbf{a},\mathbf{b})=1.0-\frac{2\left | {\mathbf{a}} \odot {\mathbf{b}} \odot {\mathbf{h}} \right |}{\left | {\mathbf{a}} \odot {\mathbf{h}} \right | + \left | {\mathbf{b}} \odot {\mathbf{h}} \right |}.
\end{align}
The losses for the refining stage can then be denoted as $\mathcal L_q^{c}{}'=\rm d'(\hat{\mathbf{y}}_q^{c}{}',\mathbf{y}_q)$, $\mathcal L_q^r{}'=\rm d'(\hat{\mathbf{y}}_q^r{}',\mathbf{y}_q)$, and $\mathcal L_q^b{}'=\rm d'(\hat{\mathbf{y}}_q^b{}', \mathbf{y}_q^b)$, respectively. In summary, the overall loss can be written as:
\begin{align}
    \mathcal L&=w\left(w^c\mathcal L_q^c+w^r\mathcal L_q^r+w^b\mathcal  L_q^b\right) \nonumber +w'\left(w^c\mathcal L_q^{c}{}'+w^r\mathcal L_q^r{}'+w^b\mathcal L_q^b{}'\right),
    \label{equ:weight}
\end{align}
where $w$ and $w'$ are the weights to balance the losses of the proposing and refining stages, and their value changes progressively during training. Especially, there are $w={\rm max}(0.5,1.0-{n}/T)$ and $w'={\rm min}(1.0,{n}/T)$, where $n$ is the current epoch number, and $T$ is an adjusting factor, forcing the model to pay more attention to optimizing the modules in the proposing stage in the early epochs of training. After the proposing stage’s convergence, the model’s focus gradually shifts to optimize the modules in the refining stage. $w^c$, $w^r$, and $w^b$ are the weights of the losses corresponding to the three prediction results from the composite module, region branch, and boundary branch in both stages. In this work, we fix them as $w^c=1.0$, $w^r=0.5$, and $w^b=0.5$.

\begin{figure}[t]
	\centering
	\includegraphics[width=\linewidth]{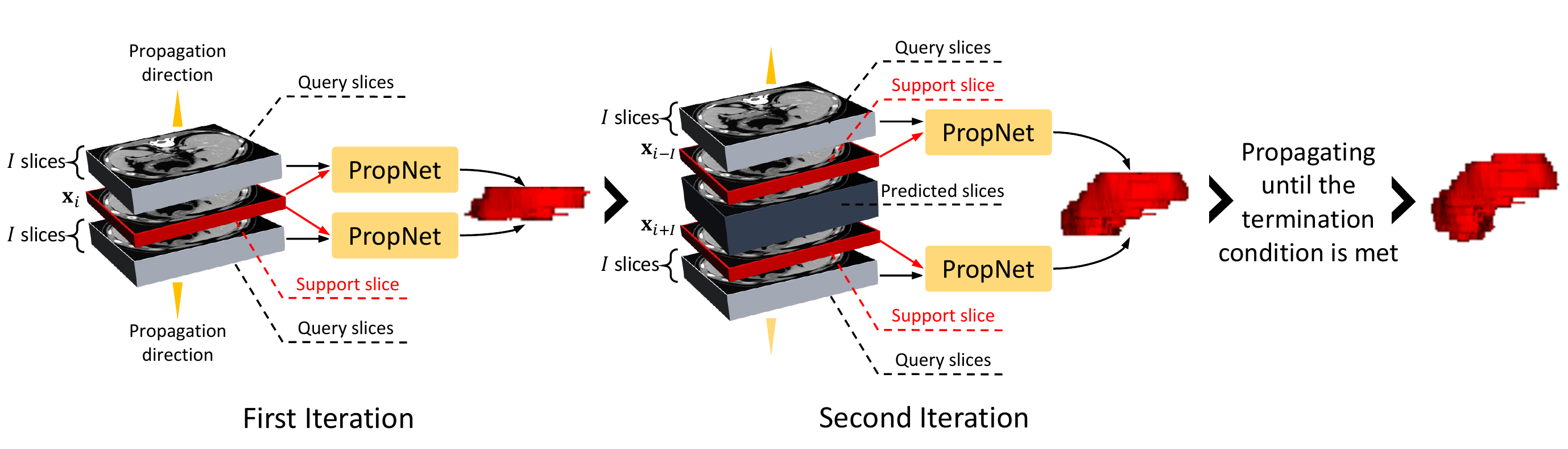}
	\caption{
	A sketch of the up-down parallel iterative inference strategy.}
	\label{fig:fig3_up_down}
\end{figure}

\subsection{Model inference}
\label{sec:infer}
Figure~\ref{fig:fig3_up_down} shows the up-down parallel iterative inference strategy. For a test 3D CT scan, we first ask the radiologists to select and annotate a transverse slice. The trained model iteratively generates the segmentation within a specific range from this slice until the complete 3D segmentation is completed. Specifically, in the first iteration, assuming that the radiologists select and annotate the $i$-th slice $\mathbf{x}_i$, the proposed model takes the $(\mathbf{x}_i,\mathbf{y}_i)$ as the initial support data. All other slices within the range of $(i-I, i+I)$ are treated as the query slices for prediction. In the second iteration, the model takes the $(i-I)$-th and $(i+I)$-th slices as support slices and further extends the segmentation $I$ slices downward and upward, respectively, etc. The iteration's termination condition is that the segmented tumor in the query image becomes less than a certain pre-defined threshold $\tau$. We also parallelize the downward and upward segmentation to improve the efficiency of model inference. After the deep neural network generates the output, we use a simple connectivity test to find the maximum connected component (MCC) in the 3D segmentation results to further reduce the small isolated false-positive responses.

\section{Experiments and Results}
\label{sec:exp_methods}

\subsection{Data Collection}
This study adheres to the principles of the Declaration of Helsinki. The CT scans are acquired using either the LightSpeed 64 VCT or the Discovery CT 750 HD scanners, with patients receiving rapid breathing training to avoid breathing artifacts. The scan parameters are set as follows: peak tube voltage of 120 kVp, automatic tube current-time product, collimation thickness of 64 x 0.625 mm, helical pitch of 0.984:1, the scanning thickness of 5-mm, and reconstructed thickness of 0.625-mm. The patients are scanned in the supine position, with the scan range extending from the top of the diaphragm to 2 cm below the lower edge of the pubic symphysis. All patients receive nonionic contrast media (3.5 ml/s, 1.5 ml/kg of body weight, iohexol 300mg I /ml, Omnipaque, GE Healthcare) and underwent arterial phase and venous phase scans 40 seconds and 70 seconds after the contrast agent injection, respectively. The Medical Ethics Committee of Beijing Cancer Hospital approve the study protocol, and all patients provide written informed consent.

We collect 232 3D CT scans of 128 patients with advanced gastric cancer who undergo neoadjuvant chemotherapy and radical resection at Peking University Cancer Hospital in Beijing, China, between 2010 and 2014. These CT scans include venous, arterial, and delayed phases. Among these patients, 95 (74\%) are male, with a median age of 62 years (range: 26-82 years). Of the patients, 98 (173 CT scans) are randomly selected for the training set, and the remaining 30 patients (59 CT scans) are used as the validation set. Two radiologists (one junior and one senior) manually annotate the 3D data, with each annotating half of the training set data to reduce the manual cost. For the validation set, the junior radiologist first annotates all data independently, and then two radiologists re-evaluate and revise the annotations to ensure the quality. Additionally, we collect 42 patients (median age of 62 years; range: 28-78 years) with stage IV gastric cancer between 2016 and 2020 at Peking University Cancer Hospital as the independent external validation set for prognostic verification. These patients receive anti-programmed cell death protein 1/programmed cell death ligand 1 (PD-1/PD-L1) antibody alone or in combination with anti-cytotoxic T lymphocyte-associated antigen 4 (CTLA-4) antibody. For the external validation set, one radiologist selects and annotates a transverse slice for each 3D CT. All 3D visualizations of the tumors in this study are created using the open-source ITK-SNAP tool~\cite{yushkevich2006user}.

\begin{table}[t]
\begin{center}
	\caption{Comparison performance of the proposed PropNet and other popular methods for segmenting gastric tumor on the validation set. The best performance is indicated in bold.
 }
	\label{table:tab1_preformance}
	\vspace*{2ex}
	\resizebox{\textwidth}{!}{
		\begin{tabular}{llccccc}
			\toprule[2pt]
			Method & Model & DSC & JI & SDSC@0.5 & SDSC@1.0 & SDSC@2.0\\
			\midrule
			Manual & Radiologist & 0.797 & 0.668 & 0.659 & 0.866 & 0.904 \\
			\midrule
			2D & UNet\cite{ronneberger2015u} & 0.700±\scriptsize{0.016} & 0.558±\scriptsize{0.015} & 0.512±\scriptsize{0.014} & 0.750±\scriptsize{0.017} & 0.810±\scriptsize{0.018} \\
			& FPN\cite{kirillov2017unified} & 0.735±\scriptsize{0.017} & 0.593±\scriptsize{0.015} & 0.535±\scriptsize{0.013} & 0.797±\scriptsize{0.017} & 0.862±\scriptsize{0.018} \\
 			& DeepLabV3\cite{chen2017rethinking} & 0.715±\scriptsize{0.020} & 0.578±\scriptsize{0.017} & 0.528±\scriptsize{0.017} & 0.772±\scriptsize{0.024} & 0.830±\scriptsize{0.025} \\
			& DeepLabV3+\cite{chen2018encoder} & 0.695±\scriptsize{0.021} & 0.549±\scriptsize{0.018} & 0.478±\scriptsize{0.016} & 0.745±\scriptsize{0.024} & 0.814±\scriptsize{0.025} \\
			\midrule
			3D & nnUNet\cite{isensee2021nnu} & 0.757±\scriptsize{0.021} & 0.636±\scriptsize{0.019} & 0.627±\scriptsize{0.017} & 0.801±\scriptsize{0.021} & 0.837±\scriptsize{0.022} \\
			& MA-MNLN\cite{zhang20213d} & 0.792±\scriptsize{0.013} & 0.662±\scriptsize{0.015} & 0.652±\scriptsize{0.012} & 0.838±\scriptsize{0.015} & 0.875±\scriptsize{0.015} \\
			\midrule
			FSL & FSS-1000\cite{li2020fss} & 0.703±\scriptsize{0.017} & 0.556±\scriptsize{0.017} & 0.568±\scriptsize{0.017} & 0.763±\scriptsize{0.015} & 0.814±\scriptsize{0.011} \\
			& SE-Net\cite{roy2020squeeze} & 0.745±\scriptsize{0.007} & 0.604±\scriptsize{0.008} & 0.605±\scriptsize{0.007} & 0.814±\scriptsize{0.007} & 0.864±\scriptsize{0.007} \\
			& GCN\cite{sun2022few} & 0.750±\scriptsize{0.011} & 0.608±\scriptsize{0.012} & 0.594±\scriptsize{0.012} & 0.818±\scriptsize{0.014} & 0.871±\scriptsize{0.012} \\
			\midrule
			& PropNet (ours) & \textbf{0.803±\scriptsize{0.015}} & \textbf{0.680±\scriptsize{0.016}} & \textbf{0.661±\scriptsize{0.016}} & \textbf{0.859±\scriptsize{0.015}} & \textbf{0.900±\scriptsize{0.014}} \\
			\bottomrule[2pt]
		\end{tabular}
	}
\end{center}
\end{table}

\subsection{Implementation Details}
For data preprocessing, we first resample the data to have a spacing of 0.6 mm/pixel in both x and y directions and cropped inputs at a resolution of $256 \times 256$, centered on the tumor in the support slice. Additionally, we experimentally normalize the CT image with a window level of 75 and a window size of 250, using the following formula: $\mathbf{x}=\left ({\mathbf{x}\odot{1}_{\{-50<\mathbf{x}<200\}}+50}\right )/{250}.$

During training, we randomly select one slice as the support slice and the adjacent four slices as query slices at each iteration. We set the erosion kernel size $E$ in equation~\ref{equ:erode} to nine pixels for generating boundary ground truth. During model inference, the radiologist first selects and annotates a slice $(\mathbf{x}_s, \mathbf{y}_s)$ as the initial support slice. Then, the proposed PropNet propagates this 2D annotation to 3D space via the up-down parallel iterative inference strategy. Note that we set the termination threshold $\tau$ of the inference strategy to $\left | \mathbf{y}_s \right |/20$, and the range parameter $I$ to $I=\max(1, \left \lfloor{20}/{spacing_z}\right \rfloor)$, where $spacing_z$ is the spacing of the z-direction in the CT scan.

We implement the proposed model using PyTorch~\cite{paszke2019pytorch} and trained it using the Adam optimizer~\cite{kingma2014adam} with an initial learning rate of 0.001. We adopt cosine annealing warm restarts~\cite{loshchilov2016sgdr} as the learning rate decay strategy, with parameters set to $T_0=40$, $T_{mult}=2$, and $\eta_{min}=5e-6$. The entire training process iterates for 200 epochs, with the adjusting factor $T$ set to 40.

For all convolutional layers, we set the stride and padding accordingly so that the resolution between the input and output features is the same unless otherwise specified (such as "ch=1" in Figure~\ref{fig:fig1_overview}). The number of output feature channels is the same as the input, and each convolutional layer is followed by a batch normalization layer and a weakly rectified linear unit (ReLU) activation layer, except for the $1\times 1$ convolutional layer. We will make our source code available on Github upon acceptance for publication.

\begin{table}[t]
\begin{center}
	\caption{Comparison of computational complexity and inference speed on one standard GPU (NVIDIA Geforce GTX 1080 Ti of 11000MHz). "-M" indicates the time of human interaction. "-I" indicates the time for model inference.}
	\label{table:tab_gflops_sct}
	\vspace*{2ex}
	\resizebox{1.0\textwidth}{!}{
		\begin{tabular}{lccccccccccc}
			\toprule[2pt]
            & Radiologist & UNet & FPN & DeepLabV3 & DeepLabV3+ & nnUNet & MA-MNLN & FSS-1000 & SE-Net & GCN & PropNet \\
            \midrule
            GFLOPs & - & 7.83 & 6.87 & 27.33 & 7.91 & - & - & 56.54 & 12.41 & 22.64 & 20.47 \\
            \midrule
            s/CT-M & 420 & & & & & & & 40 & 40 & 40 & 40 \\
            s/CT-I & & 0.397 & 0.363 & 0.575 & 0.412 & 101 & 232 & 0.852 & 0.794 & 0.795 & 0.661 \\
			\bottomrule[2pt]
		\end{tabular}
	}
\end{center}
\end{table}

\subsection{Evaluation Metrics}
We measure the computational complexity of the model using Giga Floating-point Operations Per second (GFLOPs). Additionally, we introduce the average inference time per CT (s/CT) to evaluate the model's efficiency. The overall performance of gastric tumor segmentation is evaluated using Dice Similarity Coefficient (DSC) and Jaccard Index (JI). To simulate a real-world usage scenario, all speed measurements are performed on a standard GPU (NVIDIA Geforce GTX 1080 Ti of 11000MHz). Moreover, the model's prediction performance for gastric tumor boundaries (surface) at different granularities is assessed using the surface Dice similarity coefficient (SDSC). Specifically, we use SDSC@$s$ to represent an allowable tolerance range of $s$ mm. We set $s$ to 0.5, 1.0, and 2.0 to evaluate the performance at different fine-grained levels in the experiments. We use the paired t-test for performance significance testing between models.

\begin{figure}[t]
	\centering
	\includegraphics[width=\linewidth]{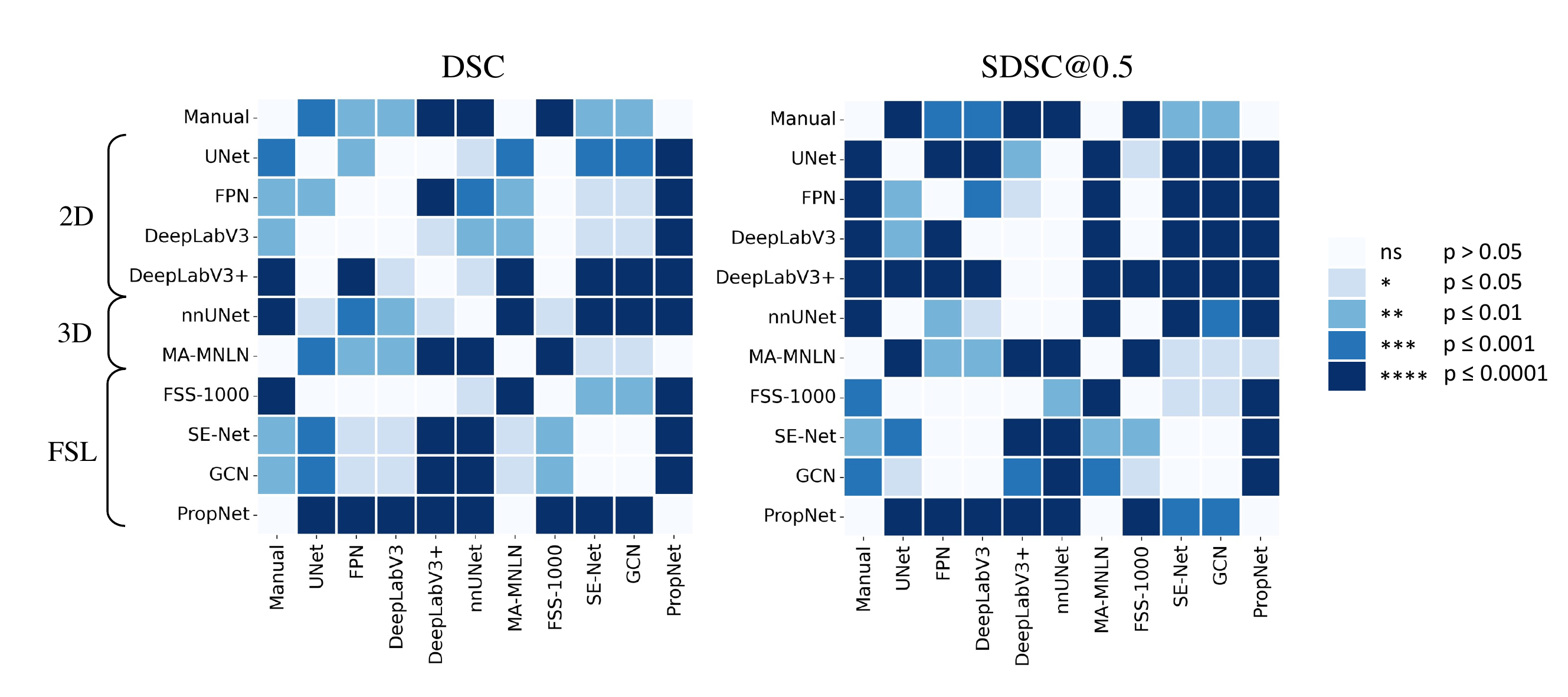}
	\caption{
	Performance significance paired t-test heatmap between different methods.}
	\label{fig:fig_p_value}
\end{figure}

\subsection{Model Performance}
To evaluate the effectiveness of our proposed PropNet, we take the repeatability between different radiologists as a manual baseline and compare our model with popular fully-supervised 2D and 3D segmentation methods. As shown in Table~\ref{table:tab1_preformance}, the inter-observer repeatability (DSC) between radiologists is 0.797. 2D methods obtain relatively low performance (between 0.695 and 0.735) due to a lack of contextual information between slices. As shown in Figure~\ref{fig:fig_p_value}, there are still significant differences between all 2D methods and the manual baseline, indicating that the performance of 2D methods was poor. Nevertheless, their network architectures depend on 2D convolutions, requiring much less computational cost and faster inference speed. For example, as shown in Table~\ref{table:tab_gflops_sct}, FPN only takes 0.363 seconds with 6.87 GFLOPs to obtain all 2D segmentations of the transverse slice in a CT scan.

In contrast, 3D methods capture overall contextual information of the volumetric data, and their accuracy is greatly improved compared to 2D methods. For example, MA-MNLN achieves a DSC of 0.792, comparable to radiologists' inter-observer repeatability with a p-value of 0.426. However, due to the limitation of GPU memory, most of the 3D methods are patch-based segmentation, which significantly increases computational costs and prolongs the inference time. Table~\ref{table:tab_gflops_sct} shows that the inference time of MA-MNLN is about 232 seconds, roughly half of the manual annotation time (420 seconds).

\begin{figure}[t]
    \begin{center}
	\includegraphics[width=\linewidth]{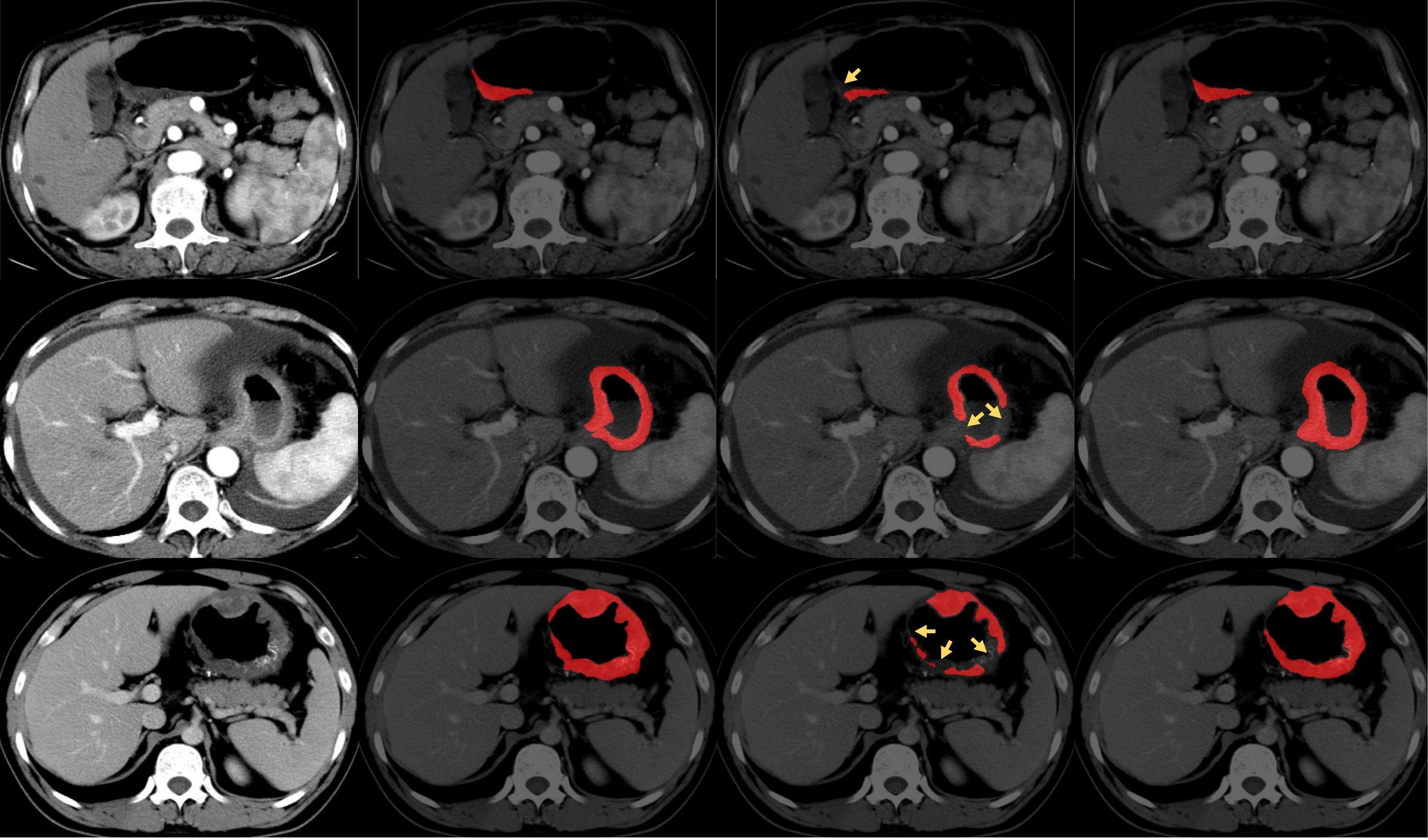}
	\caption{Visualization of two stages' predictions. From left to right are the original image, the ground truth, the proposing stage's prediction, and the refining stage's prediction. Arrows indicate the difficult-to-segment pixels in the proposing stage.}
	\label{fig:fig4_two_stage}
    \end{center}
\end{figure}

We have also implemented several few-shot learning (FSL) methods in the semi-automated scenario of the proposed PropNet to propagate an annotated 2D slice to complete 3D segmentation. These methods strike a balance between accuracy and inference speed. For instance, SE-Net can complete a CT inference in an average of 0.794 seconds (with an additional 40 seconds of human interaction) and achieves a DSC of 0.745. However, the accuracy of these methods still falls short of the level of radiologists. In contrast, our PropNet introduces specific modules to handle the complex morphologies of gastric tumors, achieving the highest DSC of 0.803 with a p-value of 0.563. It is worth noting that PropNet also achieves the highest SDSCs at all levels of surface evaluation, demonstrating its powerful ability to segment gastric tumor boundaries (surfaces). Unlike other FSL methods, our PropNet implements serialized computational operations to remove unnecessary non-local computations and applies dual-branch fusion computations to accelerate model inference on the GPU. The overall inference time for the proposed segmentation includes around 40 seconds to annotate a single transverse and an additional automated model inference time. Table~\ref{table:tab_gflops_sct} shows that the inference time for propagating a complete 3D segmentation of CT is 0.661 seconds, which is still significantly less than the time required for fully-supervised 3D methods and manual annotation. Overall, our proposed PropNet achieves comparable accuracy to the manual baseline and 3D models while also having more efficient inference efficiency and greater potential for clinical applications.

\begin{table}[t]
\begin{center}
	\caption{An ablation study to show different modules' effectiveness. "+Region" and "+Boundary" mean to add the region branch and the boundary branch, respectively. "+MCC" indicates adding the maximum connected component post-processing.}
	\label{table:tab2_ablation}
	\vspace*{2ex}
	\resizebox{0.7\textwidth}{!}{
		\begin{tabular}{lccccc}
			\toprule[2pt]
			& DSC & JI & SDSC@0.5 & SDSC@1.0 & SDSC@2.0 \\ 
			\midrule
			\multicolumn{6}{l}{\color{black} Proposing stage}\\
			+Region & 0.719 & 0.574 & 0.575 & 0.805 & 0.869 \\
			+Boundary & 0.741 & 0.600 & 0.585 & 0.817 & 0.877 \\
			\midrule
			\multicolumn{6}{l}{\color{black} Refining stage}\\
			+Region & 0.799 & 0.675 & 0.653 & 0.852 & 0.894 \\
			+Boundary & 0.800 & 0.676 & 0.657 & 0.852 & 0.893 \\
			+MCC & 0.803 & 0.680 & 0.661 & 0.859 & 0.900 \\
			\bottomrule[2pt]
		\end{tabular}
	}
\end{center}
\end{table}

We conduct several ablation experiments to evaluate the effectiveness of the proposed model. First, Table~\ref{table:tab2_ablation} shows that adding a boundary branch improves boundary segmentation, especially at more intricate levels, where the gain in accuracy at SDSC@$0.5$ is higher than at the other two levels. Secondly, introducing the refining stage improves the model's overall performance by effectively mining difficult-to-segment pixels. Figure~\ref{fig:fig4_two_stage} visualizes the predictions of the proposing and refining stages. The proposing stage fails at pixels with relatively low contrast or complex textual information, and the refining stage corrects such failures (see arrows in Figure~\ref{fig:fig4_two_stage}). Lastly, we perform an ablation test on the value of the propagation interval $I$ in model inference (Section~\ref{sec:infer}). The selection of this interval affects the model's performance and inference speed. When $I$ (in mm) is set to 5, 10, 15, and 20, DSCs are 0.798, 0.796, 0.803, and 0.797, respectively, and the corresponding inference times are 0.999, 0.759, 0.661, and 0.677 seconds, respectively. The model with an interval of 20 mm (i.e., four slices in a 5-mm CT scan) achieves the best performance and fastest inference speed, so we fix $I$ to 20 mm in other experiments. This result also shows that a medium interval restrains the possible noise caused by the excessive selection of supportive slices and maintains the information propagation between support and query slices.

\begin{figure}[t]
	\centering
	\includegraphics[width=\linewidth]{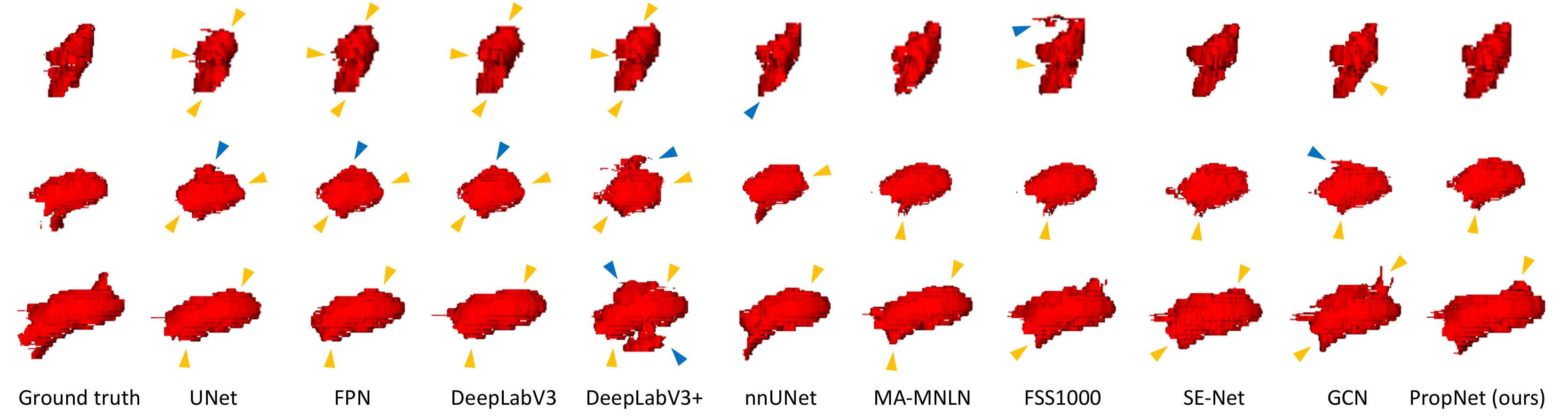}
	\caption{
		Comparison of the 3D segmentation results from three patients showing the performance of different methods. Yellow arrows indicate false-negative predictions, and blue arrows indicate false-positive predictions.
	}
	\label{fig:fig_3d}
\end{figure}

\section{Discussion}

Tumor size, as measured on resected specimens in patients with curative gastrectomy, is closely associated with tumor biological characteristics, stage, and patient prognosis~\cite{giuliani2003maximum,iwasaki2013phase,lu2013consideration}. Similarly, tumor volume measurement on CT scans could aid in assessing preoperative staging and evaluating medical response to neoadjuvant therapy~\cite{beer2006adenocarcinomas,wang2017ct}. However, in most studies, CT volumetry was conducted by manually tracing the tumor boundaries slice by slice. This process was time-consuming, and some studies have demonstrated that tumor volume segmentation took around 5-8 minutes to complete~\cite{beer2006adenocarcinomas,xu2018ct}.

Based on our literature search, there are only a few works currently focusing on automatically segmenting gastric tumors on 3D CT scans~\cite{zhang20213d,li20213d}. However, these methods require more computational costs and annotated data, which limits their application in real medical scenarios. Some works~\cite{li2020fss,roy2020squeeze,sun2022few} used few-shot learning to segment organs in 3D CT, but they do not take into account gastric tumors' irregular shape and unclear boundaries. In this pilot research report, we introduce a semi-automated method that utilizes a single 2D transverse slice annotation to infer the complete 3D gastric tumor segmentation. The proposed model (PropNet) conducts two segmentation stages: a proposing stage to segment the gastric tumors roughly and a refining stage to improve the segmentation by reducing possible minor errors.

\begin{figure}[t]
	\centering
	\includegraphics[width=\linewidth]{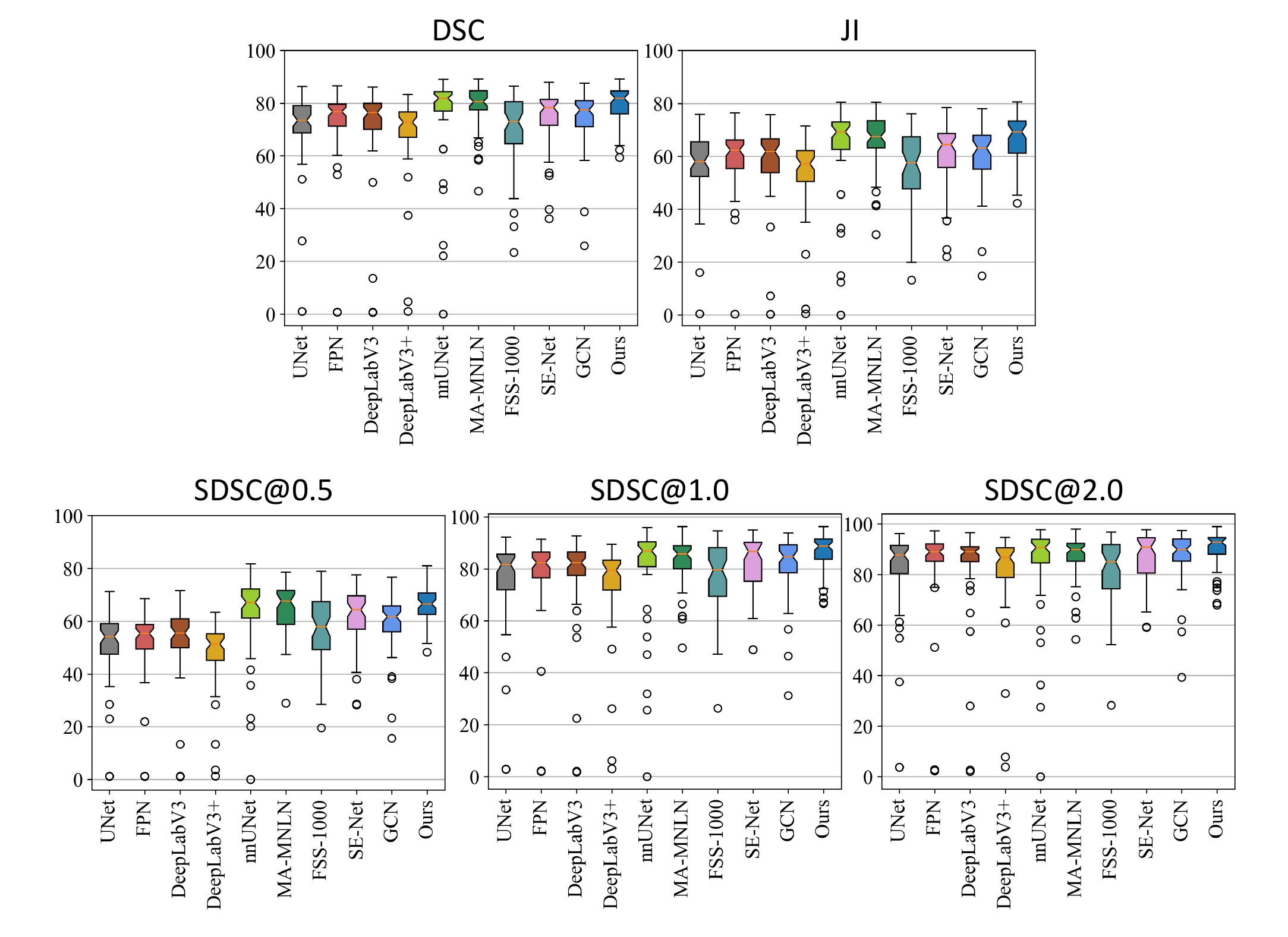}
	\caption{
	Quantitative comparison of different methods on the validation set based on box charts.}
	\label{fig:fig_comparsion}
\end{figure}

\begin{figure}[t]
	\centering
	\includegraphics[width=\linewidth]{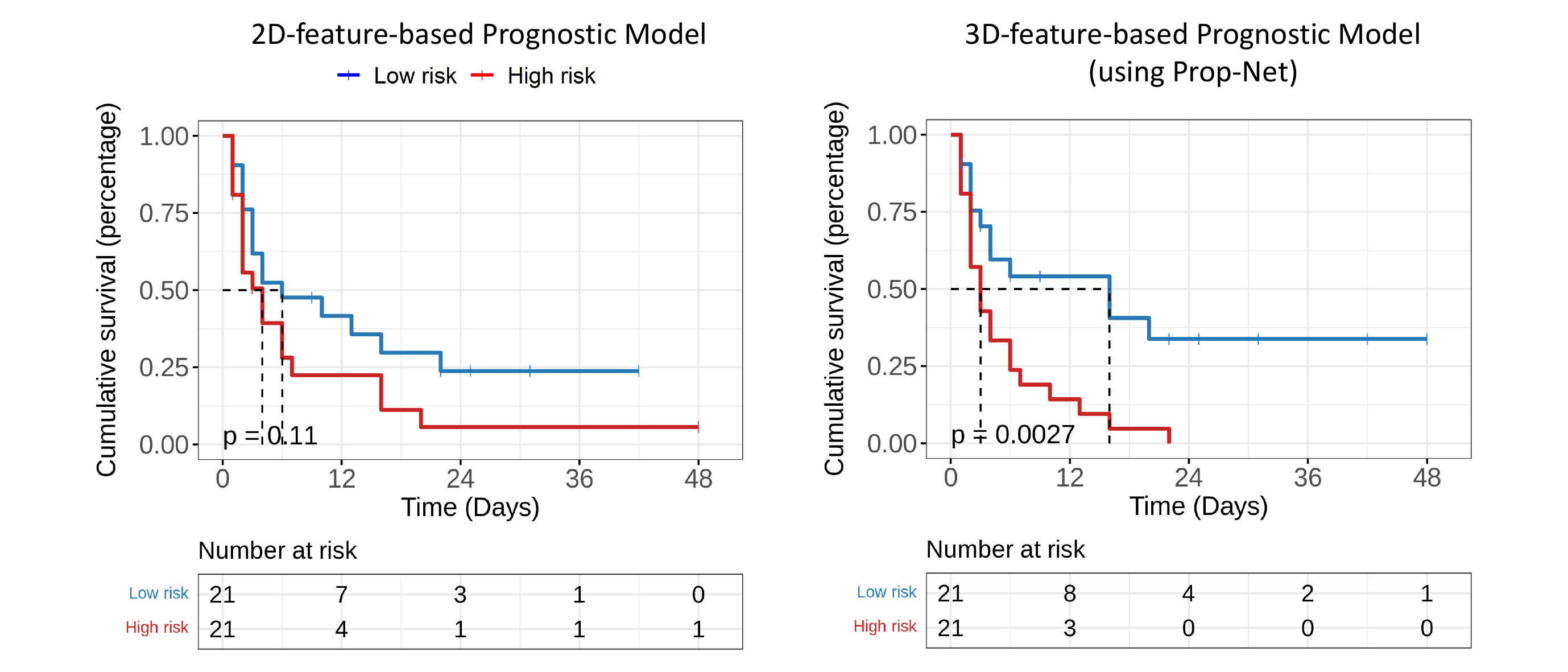}
	\caption{
		Comparison of 2D feature-based and 3D feature-based prognostic models' performance on the independent external validation set.
	}
	\label{fig:survival}
\end{figure}

We evaluate our model's performance by reproducing various advanced segmentation methods on our gastric tumor dataset. These methods include classic image segmentation methods, especially on medical images, such as U-Net~\cite{ronneberger2015u}, DeepLabV3+\cite{chen2018encoder}, nnUNet\cite{isensee2021nnu}, and MA-MNLN~\cite{zhang20213d}. We also evaluate few-shot learning-based methods, such as FSS-1000~\cite{li2020fss}, SE-Net~\cite{roy2020squeeze}, and GCN~\cite{sun2022few}. Through sufficient experimentation, we have shown the effectiveness of our method (Table~\ref{table:tab1_preformance}, Table~\ref{table:tab2_ablation}, Table~\ref{table:tab_gflops_sct}, Figure~\ref{fig:fig4_two_stage}, Figure~\ref{fig:fig_3d}, and Figure~\ref{fig:fig_comparsion}), and its potential for improving the effectiveness of downstream tasks such as survival analysis (Figure~\ref{fig:survival}).

Furthermore, Figure~\ref{fig:fig_3d} shows a comparison between different methods through 3D visualization. Classic 2D methods are limited to segmenting gastric tumors. Without the continuity of contextual information between adjacent slices, 2D methods present similar false responses (yellow or blue arrows in Figure~\ref{fig:fig_3d}). Although the proposed method uses a 2D architecture, the inter-slice context module recognizes the relationship between slices, enabling the model to efficiently extract 3D information while maintaining spatial continuity of segmentation results.

Additionally, the low computational cost and efficient inference make the model suitable for deployment in clinical scenarios. Our experiments demonstrate that the proposed method can provide human-level delineation of gastric tumors on 3D CT scans in less than one second using a standard GPU (NVIDIA GeForce GTX 1080Ti) and within 70 seconds using the commonly used CPU (Intel CORE i5-7200U with 2.5GHz).

\begin{table}[t]
\begin{center}
	\caption{The performance of our proposed method varies with different deviations of the selected and annotated initial support slice. "-" indicates an upward deviation in the CT scan, and "+" indicates a downward deviation.}
	\label{table:interval}
	\vspace*{2ex}
	\resizebox{0.7\textwidth}{!}{
		\begin{tabular}{cccccc}
			\toprule[2pt]
			Deviation (mm) & DSC & JI & SDSC@0.5 & SDSC@1.0 & SDSC@2.0 \\
			\midrule
			-15 & 0.790 & 0.665 & 0.648 & 0.842 & 0.884 \\
			-10 & 0.790 & 0.664 & 0.648 & 0.843 & 0.886 \\
 			-5 & 0.793 & 0.668 & 0.653 & 0.849 & 0.890 \\		
 			0 & \textbf{0.803} & \textbf{0.680} & \textbf{0.661} & \textbf{0.859} & \textbf{0.900} \\
			+5 & 0.798 & 0.674 & 0.655 & 0.847 & 0.887 \\
			+10 & 0.790 & 0.665 & 0.652 & 0.852 & 0.893 \\
			+15 & 0.785 & 0.662 & 0.648 & 0.838 & 0.878 \\
			\bottomrule[2pt]
		\end{tabular}
	}
\end{center}
\end{table}

Due to the complex morphologies and contexts of gastric tumors and the differences in clinical training and experience, the initial support slice's annotation can vary in two aspects: 1) different annotated radiologists; 2) different selections of the support slice. To account for this real-world scenario, we conduct experiments to evaluate the performance of the proposed PropNet when guided by different radiologists. The experimental results show that PropNet can still achieve comparable results (DSC of 0.797, JI of 0.657, SDSC@0.5 of 0.639, SDSC@1.0 of 0.851, and SDSC@0.896) when using the initial support slice chosen by another radiologist. Additionally, we perform a robustness check on the initialization of our method. Table~\ref{table:interval} shows that PropNet's performance is relatively stable, with a DSC of 0.785 even when the initialization is 15 mm away from the largest cross-section, outperforming most previous methods.

Survival analysis based on radiomics can further evaluate the clinical value of gastric tumor segmentation on CT scans. For example, Jiang et al.~\cite{jiang2020noninvasive} extracted radiomic features from manually segmented gastric tumors to predict the abundance of micro-level biomarkers. To assess the prognostic performance of our proposed method, we collect CT scans and progression-free survival (PFS) data from 42 gastric cancer patients from Beijing Cancer Hospital as the independent validation set. The radiologists select and annotate the largest slice of the tumor. We conduct two experiments based on these data and annotations.

In the first experiment, we extract 58 features, including 10 2D-shape features and 19 first-order statistics of venous and arterial phases, respectively, using only the 2D annotated largest slice. LASSO regression with four-fold cross-validation is used to select the five most important radiomic features. Based on these selected radiomic features, we use Leave-One-Out Cross-Validation (LOOCV) to evaluate the prognostic performance of a Cox proportional hazards regression model. The Harrell's concordance index (c-index) of the 2D baseline model is 0.576 (95\% CI: 0.477-0.675) (hazard ratio is 1.767 with a p-value of 0.11).

In the second experiment, we use our proposed model to propagate 3D gastric tumor segmentation from the annotated slice. We then extract 70 features, including 16 3D-shape features and 19 first-order statistics of venous and arterial phases, respectively. This 3D prognostic analysis model achieves a c-index of 0.620 (95\% CI: 0.523-0.718) (hazard ratio is 2.883 with a p-value of 0.0027), which can better distinguish high-risk and low-risk groups than the 2D baseline (see figure~\ref{fig:survival}). These experiments demonstrate that the 3D segmentation generated by our proposed PropNet can provide valuable information for downstream research, such as the survival analysis of gastric cancer patients.

There are several limitations of the proposed method that require further investigation. Firstly, the number of subjects in our study is relatively small, and we are currently conducting research on a larger cohort of gastric cancer patients who have undergone different types of interventions. Secondly, the proposed model requires manual annotation of the tumor in a 2D transverse slice by a radiologist. To improve the initialization of the model, we are considering reducing the workload on the annotation process by using extreme points or bounding boxes instead of a complete delineation of the tumor boundary in 2D. Thirdly, we aim to investigate the relationship between the quantitative statistics derived from our segmentation and the biomarkers that are useful for the prognostic analysis of gastric tumors.

\section{Conclusions}
To segment 3D gastric tumors with irregular shapes and blurred boundaries, we propose a semi-automated propagating model called PropNet that utilizes the relationship between adjacent transverse slices to propagate complete gastric tumor segmentation from the annotation of a 2D slice. We design a series of modules to improve the gastric tumor segmentation, including adding a boundary branch to enhance model performance at tumor boundaries, using a refining stage to correct possible errors for difficult-to-segment pixels, and proposing an up-down parallel strategy to achieve higher efficiency for the inference of test CT scans. Experimental results show that the proposed PropNet can achieve human-level segmentation performance and significantly improve segmentation efficiency. We believe this novel segmentation framework has the potential to improve the quantitative analysis of gastric tumors, particularly for tasks with high-throughput image reading.

\section{Acknowledgments}
This work is supported in part by the Beijing Natural Science Foundation under Grants Z200015 and Z180001, the National Natural Science Foundation of China under Grants 12090022, 11831002, 91959205, 81801778, PKU-Baidu Foundation under Grant 2020BD027, Pilot Project of Public Welfare Development Reform of Beijing-based Medical Research Institutes under Grant 2019-1.

\section{Conflict of Interest}
The authors have no conflict to disclose.

\clearpage

\section*{References}
\addcontentsline{toc}{section}{\numberline{}References}
\vspace*{-20mm}

\bibliographystyle{medphy}    
\bibliography{main}


\end{document}